\pdfoutput=1

\documentclass[twocolumn]{aastex631}
\usepackage{epstopdf}
\usepackage{xspace}
\usepackage{graphicx}
\usepackage{amssymb}
\usepackage{amsmath}
\usepackage{natbib}
\usepackage{float}
\usepackage{hyperref}

\shorttitle{Microlensing Events in the Lightcurve of S5\,0716+714}
\shortauthors{Kr\'{o}l et al.}

\begin{document}

\title{Possible Gravitational Microlensing Events in the Optical Lightcurve of Active Galaxy S5\,0716+714}

\correspondingauthor{D.~{\L}.~Kr\'{o}l}
\email{dominika.l.krol@doctoral.uj.edu.pl}

\author[0000-0002-3626-5831]{D.~{\L}.~Kr\'{o}l}
\affiliation{Astronomical Observatory of the Jagiellonian University, Orla 171, 30-244 Krak\'{o}w, Poland}

\author[0000-0001-8294-9479]{{\L}.~Stawarz}
\affiliation{Astronomical Observatory of the Jagiellonian University, Orla 171, 30-244 Krak\'{o}w, Poland}

\author[0000-0001-8320-3919]{J.~Krzesinski}
\affiliation{Astronomical Observatory of the Jagiellonian University, Orla 171, 30-244 Krak\'{o}w, Poland}

\author[0000-0002-4377-0174]{C.~C.~Cheung}
\affiliation{Naval Research Laboratory, Space Science Division, Washington, DC 20375, USA}

\begin{abstract}

A well-known active galaxy of the blazar type, S5\,0716+714, is characterized by a particularly high variability duty cycle on short-time scales at optical frequencies. As such, the source was subjected to numerous monitoring programs, including both ground-based as well as space-borne telescopes. On closer inspection of the most recent accumulation of the data provided by the Transiting Exoplanet Survey Satellite, we have noticed several conspicuous events with `volcano-like' symmetric shape, lasting all for several hours, which closely resemble the achromatic events detected with the previous Whole Earth Blazar Telescope campaigns targeting the source. We propose that those peculiar features could be due to the gravitational micro-lensing of the innermost segments of the precessing jet in the system, by a binary lens. We study the magnification pattern of the lens with the inverse-ray shooting method, and the source trajectory parameters with the Python package \textsc{MuLensModel}. In this way, we were able to fit successfully all the selected events with a single lens, adjusting slightly only the source trajectory parameters for each lensing event. Based on the fit results, we postulate the presence of a massive binary lens, containing an intermediate-mass black hole, possibly even a super-massive one, and a much less massive companion (by a factor of $\lesssim 0.01$), located within the host galaxy of the blazar, most likely the central kiloparsec region. We discuss the major physical implications of the proposed scenario regarding the quest for the intermediate-mass and dual supermassive black holes in active galaxies.

\end{abstract}

\section{Introduction}
\label{sec:intro}

Gravitational lensing has wide application in cosmology and astrophysics \citep[for reviews see, e.g.,][]{Blandford92,Narayan96}. Among many others, it was proposed that microlensing can possibly explain at least some part of the variability of active galactic nuclei (AGN), in particular the achromatic very short-time scale flux changes in blazar sources (e.g., \citealt{Subramanian85}, \citealt{Schneider87}, \citealt{Gopal91}; for a review see \citealt{Wambsganss06}). Lately, this idea has been somewhat resurrected by \cite{Vedantham17} and \cite{Peirson22}, who proposed that the peculiar year-long features in the radio lightcurve of the blazar J1415+1320, consisting of a symmetric double-horn minimum (designated by the authors as a `volcano-type' shape event), could be explain by `milli-lensing' events involving a binary lens. 

The two-point-mass lensing is a well-understood phenomenon \citep{Schneider86}. For example, gravitational lensing by a binary star leading to the double-horn shape in the lightcurve of a lensed source, has been discussed by \cite{Mao91} in a specific context of the Galactic stars and planets. The authors pointed out that, when the source is projected close to the caustics, an extra pair of unresolved micro-images is formed, creating complicated patterns in the source light curves, including the characteristic double-horn features.

In this paper, we present another take on the story. Namely, we identify peculiar symmetric features in the optical lightcurve of the blazar S5\,0716+714, lasting for several hours each, with similar double-horn flux profiles. We propose that all of these could in fact be due to the lensing of the innermost segments of the precessing jet in the source by a binary lens, consisting of an intermediate-mass black hole (IMBH), or even a supermassive mass black hole (SMBH), with a less-massive companion (mass ratio $\lesssim 0.01$), located within the host galaxy of the blazar.

S5\,0716+714 is a very well-known blazar located at  the redshift $0.227 < z < 0.254$ \citep{Dorigo22}. For the modern cosmology with $H_0 = 70$\,km\,s$^{-1}$\,Mpc$^{-1}$, $\Omega_{\rm m} = 0.3$, and $\Omega_{\Lambda} = 0.7$, this corresponds to the angular-size distance of $d_{\rm A} \simeq (0.75-0.82)$\,Gpc, and the conversion scale of $\sim (3.64-3.96)$\,kpc\,arcsec$^{-1}$. The blazar is characterized by a particularly high duty cycle on short time scales at optical frequencies, and as such was subjected to numerous intra-night monitoring programs, including several campaigns on behalf of the Whole Earth Blazar Telescope (WEBT) Collaboration\footnote{\url{http://www.oato.inaf.it/blazars/webt}} \citep[e.g.,][]{Ostorero06,Bhatta16}, and most recently with the Transiting Exoplanet Survey Satellite ({\em TESS}) by \citet{Raiteri21}.

On closer inspection of the most recent {\em TESS} light curves of the target, we have noticed at least six several-hour-long conspicuous events, whose `volcano-like' symmetric shape resembled the features discussed by \citet{Vedantham17} and \citet{Peirson22} in the radio lightcurve of J1415+1320. We note however that the J1415+1320 radio events are much longer, lasting up to even a few years. Interestingly, the identified {\em TESS} events resemble closely achromatic features detected previously in the source optical lightcurve with WEBT \citep{Bhatta16}. As argued below, all of those could be explained in the framework of the microlensing scenario.

\begin{figure}[th!]
    \centering
    \includegraphics[width=\columnwidth]{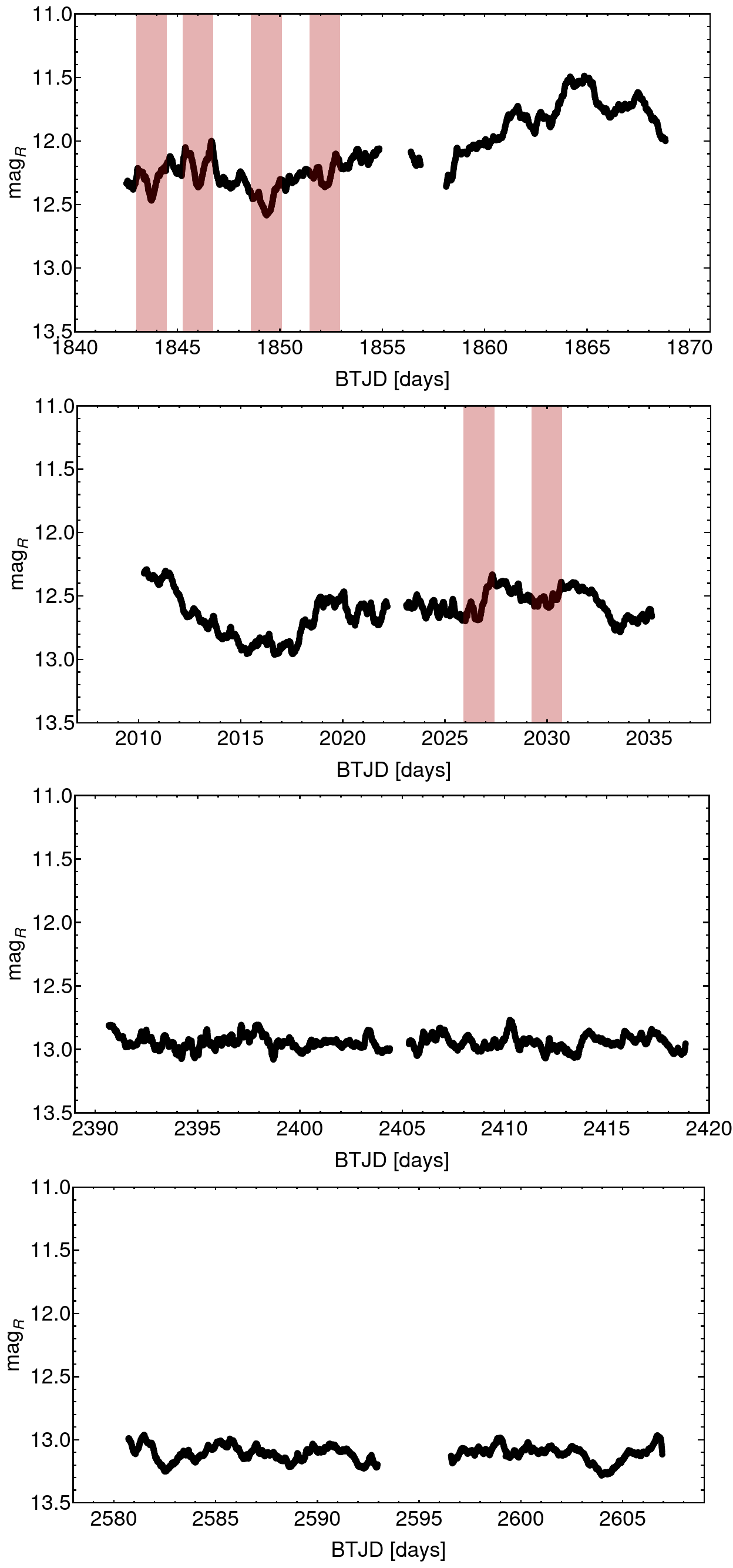}
   \caption{Four month-long {\em TESS} observations of S5\,0716+714 investigated in this paper. The {\em TESS} data were converted to R-band magnitudes and binned in 10-min intervals (see Section\,\ref{sec:TESS}).}
    \label{fig:TESS}
\end{figure}

\section{{\em TESS} Data}
\label{sec:TESS}

For our analysis, we use short-cadence (120 sec) {\em TESS} mission \citep{Ricker14} data for S5\,0716+714 ({\em TESS} Input Catalog source, TIC 147796121) downloaded from the Barbara A. Mikulski Archive for Space Telescopes (MAST)\footnote{https://archive.stsci.edu/index.html}. The target was observed in four seasons:
\begin{itemize}
\setlength\itemsep{0.0em}
\item[1)] 2019-12-25 -- 2020-01-20 sector\,20, camera\,2, ccd\,2; 
\item[2)] 2020-06-09 -- 2020-07-04  sector\,26, camera\,4, ccd\,4; 
\item[3)] 2021-06-25 -- 2021-07-23 sector\,40, camera\,4, ccd\,3;
\item[4)] 2022-01-01 -- 2022-01-27 sector\,47, camera\,2, ccd\,2;
\end{itemize}
corresponding to the four panels of Figure\,\ref{fig:TESS}. We adopted BTJD$=$BJD$-$2457000. Only the first observational season was the subject of the detailed analysis by \citet{Raiteri21}.

The Target Pixel Files data (containing the original CCD time series data in shape of postage stamp image cutouts) from all the seasons were checked for possible field stars, which could fall within the apertures/masks used by {\em TESS} Science Processing Operations Center pipeline ({\em TESS}-pipeline) to extract the light curves.

\begin{figure}[th!]
    \centering
    \includegraphics[width=\columnwidth]{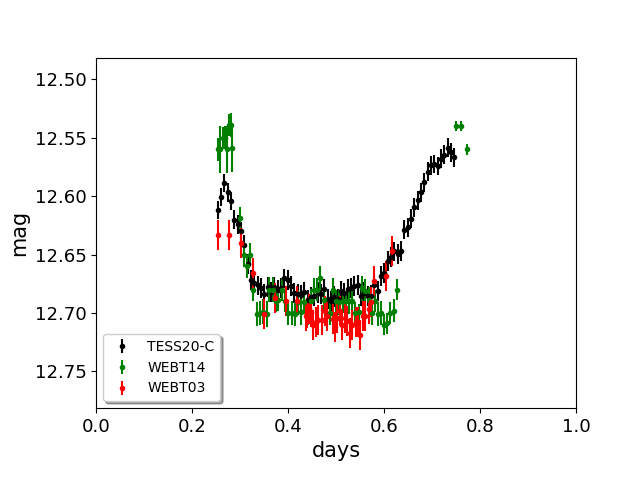}
   \caption{Zooms to the {\em TESS} event on MJD 9026 (black data points), with the superimposed data for the plateau events present in the 2003 WEBT data \citep[red symbols; from][shifted vertically by --1.45\,mag]{Ostorero06}, and 2014 WEBT data \citep[green symbols; from][shifted vertically by --1.3\,mag]{Bhatta16}. The WEBT datapoints were taken directly from Figure\,9 in \citet{Bhatta16}.}
    \label{fig:zooms}
\end{figure}

S5\,0716+714 appears to be a well isolated object and only faint ($>19$\,mag) field stars can be found in the {\em TESS}-pipeline masks. This is confirmed by the CROWDSAP factor (describing the amount of light in the aperture attributed to the target object), which is between 0.962--0.982 in all observing seasons. A small, but negligible, light contamination comes from neighboring bright stars outside of the {\em TESS}-pipeline masks. Therefore, for the further analysis we used Pre-search Data Conditioning SAP fluxes (systematics corrected photometry using co-trending basis vectors), which are the light curves prepared by the {\em TESS}-pipeline. The {\em TESS} fluxes $F$ were converted to R-band magnitudes following \citet{Raiteri21}, namely ${\rm mag_{TESS}} = - 2.5 \log F + 20.42$. For all subsequent analysis, we binned the data in 10-min intervals. The resulting lightcurves of S5\,0716+714 are presented in the four panels of Figure\,\ref{fig:TESS}. 

After a close inspection of the acquired {\em TESS} lightcurves, we found six notable events displaying symmetric double-horn minima. In fact, these lightcurve shapes resemble events previously found in ground-based WEBT campaigns in 2003 \citep{Ostorero06} and 2014 \citep{Bhatta16}, where {\it ``for about 6hr (...) the source suddenly exhibited a strongly reduced level of flux variability, resulting in a plateau in all four bands’ light curves''} (\citeauthor{Bhatta16}, Section\,3.1.4 therein). One of the selected {\em TESS} events, with the superimposed WEBT 2003 and 2014 events (shifted vertically by --1.45\,mag and --1.3\,mag, respectively), is presented in Figure\,\ref{fig:zooms}. Note that the offsets of $\sim 1.3$\,mag, translating to a factor of $\sim 3$ in flux, implies that the double-horn events can be found at a relatively wide range of a flux level of the source. Moreover, as we argue below, such a repeating pattern does not conform with the stochastic, red-noise type variability of the blazar \citep[e.g.,][]{Rani13,Bhatta16,Raiteri21}.

We emphasize the achromatic character of the 2014 WEBT plateau event, that was clearly demonstrated by \citet{Bhatta16}. As for the {\em TESS} data, while here we show only the {\em TESS} light curve converted to R-band magnitudes, the accompanying WEBT monitoring --- even though with a more sparse coverage and not overlapping with the entirety of the {\em TESS} pointings --- confirms the achromatic character of the flux changes also in the 2019/20 dataset \citep[see Figure\,3 in][]{Raiteri21}. In particular, four out of the six events selected here for the analysis, have the accompanying WEBT data, which clearly show that the flux changes traced by the dense {\em TESS} sampling, are followed closely in all the B, V, R, and I filters separately.

All in all, during the total of four months of the {\em TESS} monitoring time, we identified six potential microlensing events in the optical lightcurve of S5\,0716+714. The question therefore arises on the frequency and the duty cycle of such events. In this respect, first we note that there may be more of similar features in the {\em TESS} lightcurve, which are however more difficult to spot because of the superimposed flaring activity of the jet, contaminating the magnification pattern. And indeed, due this reason, several other possible features in the {\em TESS} data were not selected for the analysis presented below: when viewed with a sufficient zoom, they either appeared asymmetric, or the plateau during the flux minimum phase displayed some small-amplitude variability. Second, we note that, whenever an intense monitoring campaign with sufficiently long time coverage (of at least a handful of days) and sufficiently dense sampling (like the WEBT 2003/2014 or the {\em TESS} campaigns) had targeted the source, at least one microlensing candidate feature could be seen. On the other hand, the majority of the intra-night observations of the blazar with a single ground-base telescope, lasting for several hours per night at most, are typically insufficient to cover the entirety of a single magnificiation event, precluding any robust identification of such.

\begin{deluxetable*}{ccccccc}[t!]
\tablecaption{\textsc{MulensModel} best-fit parameters. \label{tab:results}}
\tablehead{\colhead{Parameter} & \colhead{TESS19-A} & \colhead{TESS19-B} & \colhead{TESS20-A} & \colhead{TESS20-B} & \colhead{TESS20-C}  & \colhead{TESS20-D} }
\startdata
$\rho$	&	0.013$^f$	&	0.013$^f$	&	0.013$^f$	&	0.013$^f$	&	0.013$^f$	&	0.013$^f$ \\
$q$	&	0.00545$^f$	&	0.00545$^f$	&	0.00545$^f$	&	0.00545$^f$	&	0.00545$^f$ &	0.00545$^f$\\
$d$	&	0.727$^f$	&	0.727$^f$	&	0.727$^f$	&	0.727$^f$	&	0.727$^f$	&	0.727$^f$ \\
$u_0$	& 0.36	&	0.35	&	0.48	&	0.78	&	0.75	&	0.54 \\
$t_E$	& 3.0	&	4.0	&	3.4	&	2.4	&	1.5	&	1.7 \\
$\alpha$	& 89	&	94	&	82 &	88	&	87	&	94 \\
$t_0$ [BTJD]	& 1843.78	&	1845.92	&	1849.62 &	1852.30	&	2026.66	&	2030.00  \\
\hline
$t_0$ [UTC]	& 2019-12-26 & 2019-12-28 & 2020-01-01 & 2020-01-03 & 2020-06-26 & 2020-06-29 \\
	        &  06:36  &  09:58  &  02:46  &  19:05  &  03:55  &  12:04  \\
$\chi^2$/dof	& 1.91	& 1.5	& 3.75	& 2.10	& 0.93	& 0.89 \\	
\enddata
\tablecomments{ $^f$ parameter frozen in the fitting procedure.}
\end{deluxetable*}

To sum up, we conclude that the symmetric double-horn features in the optical lightcurve of S5\,0716+714, with the peak-to -peak duration of about 12\,hr and the flux minimum phase lasting for a few/several hours, are promising microlensing event candidates. Those events appear relatively frequent, although one needs more good-quality monitoring data to estimate robustly the duty cycle and clustering properties of such.

\section{Binary Lens Modeling}
\label{sec:model}

In the first stage of our analysis, we used the inverse-ray shooting method to obtain magnification patterns for different binary parameters. If we consider objects with masses $M$ and $q M$, where $q<1$, at the respective positions ${\mathbf{x}}_M=[0,0]$ and ${\mathbf{x}}_{qM}=[0,d]$ on the lens plane, the lens equation takes the form
\begin{equation}
    {\mathbf{y}}={\mathbf{x}}-\frac{{\mathbf{x}}}{|{\mathbf{x}}^2|} - q \, \frac{{\mathbf{x}}-{\mathbf{d}}}{|{\mathbf{x}}-{\mathbf{d}}|^2} ,
    \label{eq:lens}
\end{equation}
where ${\mathbf{y}}$ and ${\mathbf{x}}$ are two-dimensional angular coordinates of the rays on the source and lens planes \citep{Schneide92}. In the above, all the quantities are in units of the Einstein angle
\begin{equation}
\Theta_E =\left(\frac{4GM}{c^2}\;\frac{D_{\rm LS}}{D_{\rm L} D_{\rm S}}\right)^{1/2} ,
    \label{eq:angle}
\end{equation}
where $D_{\rm S}$, $D_{\rm L}$, and $D_{\rm LS}$ are the angular distances to the source, to the lens, and between the source and the lens, respectively. Note that the Einstein radius $R_E = \Theta_E D_{\rm L}$.

To obtain magnification patterns of the lens characterized by given $q$ and $d$ in the inverse ray shooting method, we evenly distributed a set of rays on the lens plane and calculated the position of each ray on the source plane. For each bin of the rays, the magnification is the ratio between the number of rays which ended up in it in the presence of, and in the lack of the lens. In our investigation of the parameter space, we have calculated magnification patterns for various binary mass ratios from the range $q \in [0.001,0.5]$ and separations $d \in [0.5,1.5]$. We note that for small separations $d$, magnification occurs predominantly in a compact area surrounding the centre of mass of a binary lens. If, in addition, the mass ratio is small (in practice, below 0.001), the magnification pattern resembles a point-mass lens. As the mass ratio increases, the area of magnification is also enlarged, creating various shapes which exact appearance depends on the separation parameter. The vulcano-type events with durations and magnifications resembling the ones presented in Figure\,\ref{fig:zooms}, could be obtained for the combination $q \lesssim 0.01$ and $d \simeq 0.7$.

\begin{figure*}[th!]
    \centering
    \includegraphics[width=0.49\textwidth]{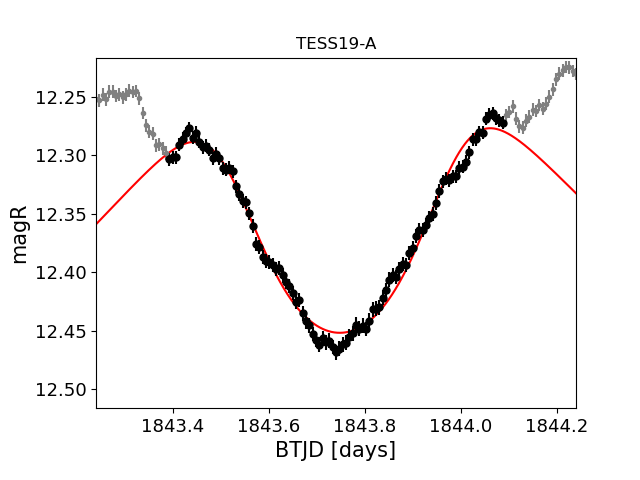}
        \includegraphics[width=0.49\textwidth]{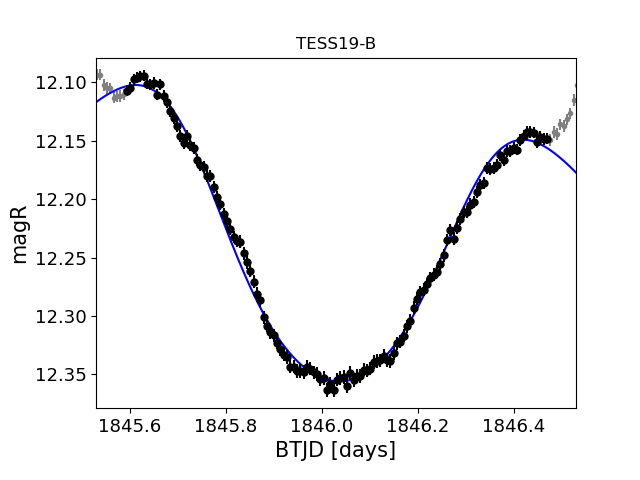}
            \includegraphics[width=0.49\textwidth]{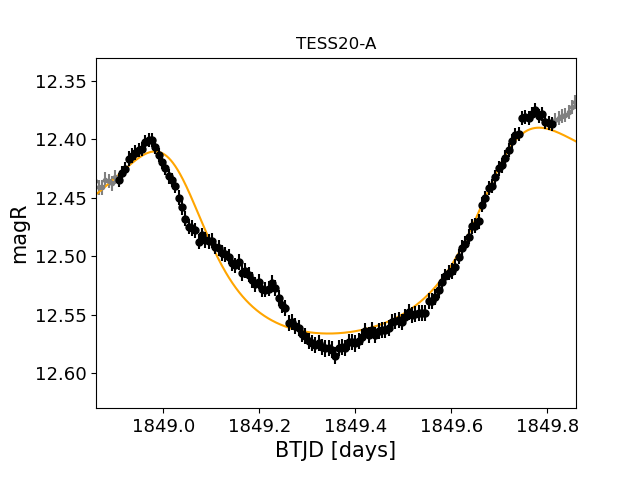}
                \includegraphics[width=0.49\textwidth]{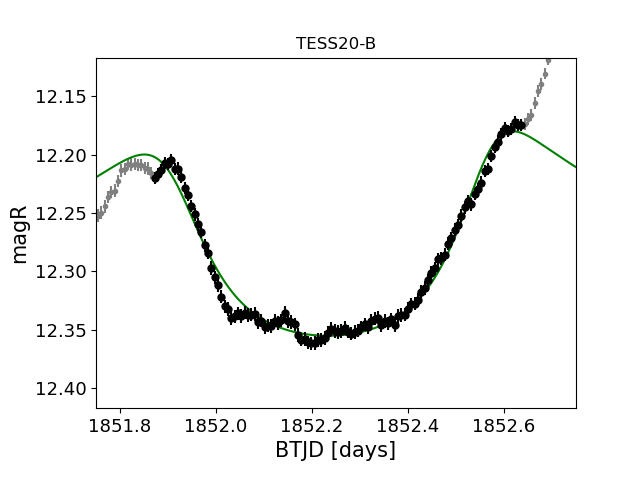}
                    \includegraphics[width=0.49\textwidth]{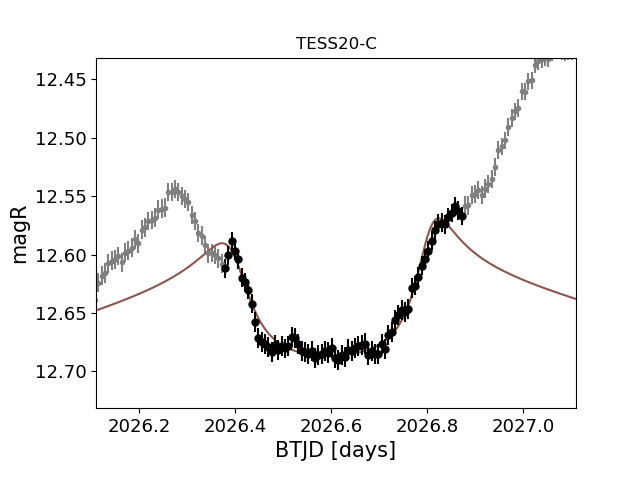}
                        \includegraphics[width=0.49\textwidth]{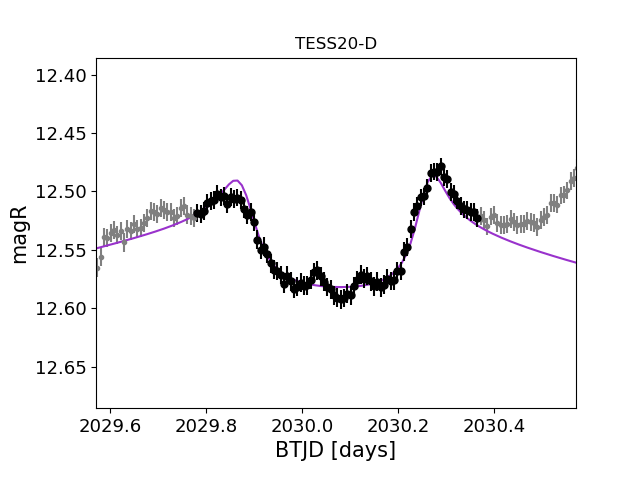}
   \caption{The \textsc{MulensModel} fits (as described in Section\,\ref{sec:model} and summarized in Table\,\ref{tab:results}) to the six selected events in the {\em TESS} light curve of S5\,0716+714 (see Figure\,\ref{fig:TESS}).}
    \label{fig:results}
\end{figure*}

The binary parameters, $q$ and $d$, determine the magnification pattern. However, in order to obtain a specific magnification curve matching a given feature present in the lightcurve, we need four parameters defining the source trajectory (the impact parameter between the source and the center of the mass of the lens in units of the Einstein angle, $u_0$; the Einstein crossing time, $t_E$; the time of the closest approach between the source and the lens, $t_0$; and the angle between the source trajectory and binary lens axis, $\alpha$), as well as the source parameter $\rho$, which is the radius of the source expressed as a fraction of the Einstein radius. In the next step of the analysis, we have therefore performed a formal fitting of those parameters to the six selected events of the source light curve, utilizing the Python package \textsc{MuLensModel} for modeling microlensing \citep{Poleski18}. For magnification calculation we have chosen the VBBL method \citep{Bozza10,Bozza18} implemented in \textsc{MuLensModel}; fitting was performed by minimalizing $\chi^2$, using the algorithm described in \cite{Gao12}.

\begin{figure*}[th!]
    \centering
    \includegraphics[width=0.42\textwidth]{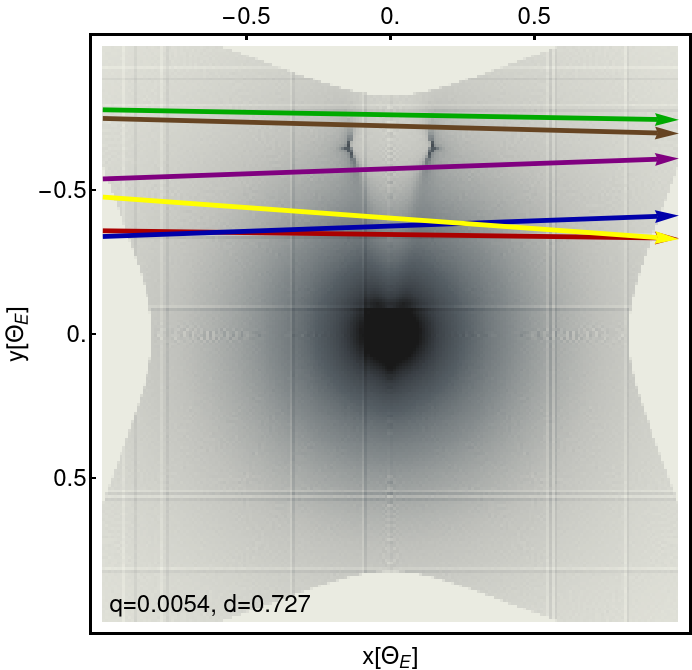}
    \includegraphics[width=0.56\textwidth]{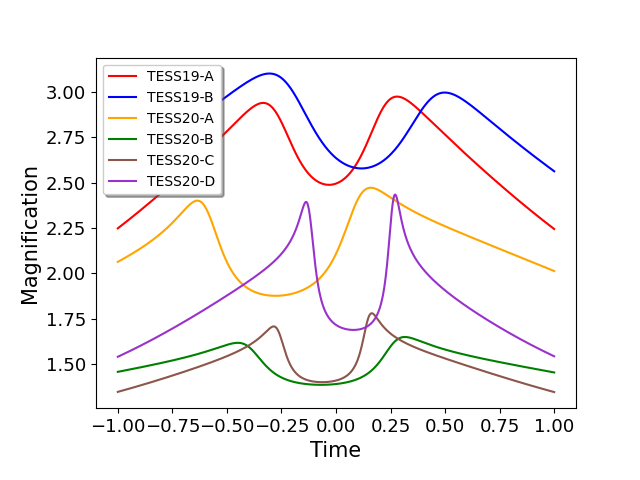}
   \caption{The magnification pattern along with the source trajectory (left panel), and the resulting source light curves (right panel), corresponding to the six \textsc{MulensModel} fits presented in Figure\,\ref{fig:results}.}
    \label{fig:map}
\end{figure*}

Due to a large number of parameters, it was impossible to perform fitting with all of the model parameters allowed to vary at the same time. For this reason, keeping in mind the general features of the explored magnification patterns from the inverse ray-shooting studies, in the first step of the analysis with the \textsc{MuLensModel}, we investigated the lens parameters by fixing $q$ and $d$ at different values for various source trajectories, and compared the quality of the obtained fits. It was relatively straightforward to find in this way lenses that allowed to reproduce either the first two (TESS19-A and TESS19-B), or the last four (TESS20-A to TESS20-D) events; however, finding a lens that can fit all six events for similar source trajectories was challenging, and the best results could be obtained for the $q\sim 0.005$ and $d\sim 0.75$ combination. To derive the final, more precise values of the lens parameters, in the next step of the analysis, we fit the TESS20-D event for a given (fixed) representative source trajectory, with $q$ and $d$ allowed to vary (initial values of 0.01 and 0.7, respectively), resulting in the best-fit values of $q=0.00545$ and $d=0.0727$. In the final, third step of the fitting procedure, the fit was performed for all the selected events with the fixed lens parameters $q=0.00545$ and $d=0.0727$, by thawing the source trajectory parameters $t_E$, $t_0$, $u_0$, and $\alpha$. It should be noted at this point, that the particular choice of the $\rho$ parameter does not affect the \textsc{MuLensModel} fitting results, as long as the source appears point-like, meaning $\rho \ll 1$. The final model was chosen based on the $\chi^2$ statistic.

As shown on Figure\,\ref{fig:results} and summarized in Table\,\ref{tab:results}, the selected events could all be fitted reasonably well in this way with only slight changes in the source trajectory parameters $u_0$, $t_E$, $\alpha$, and $t_0$. This is further visualized in Figure\,\ref{fig:map}, where we present the magnification map of the lens along with the source trajectory, along with the resulting source light curves, corresponding to the  \textsc{MulensModel}. The emerging set of the source trajectory (on the lens plane) parameters is obviously not unique, and it is possible that comparably good fits could be obtained for the other values of $u_0$, $t_E$, $\alpha$, and $t_0$. The point of the fitting exercise presented here is, however, to check whether \emph{the same single binary lens} is sufficient to explain all the selected events. Moreover, we claim that this lens has to be characterized by the parameters close to selected values of $q \lesssim 0.01$ and $d \sim 0.7$.

\section{Discussion and Conclusions}
\label{sec:discussion}

The distance to the active galaxy S5\,0716+714, $z=0.31\pm0.08$, has been determined by \citet{Nilsson08}, based on the ``standard candle'' assumption for the detected host galaxy. \citet{Danforth13}, based on the detection of the narrow Ly$\alpha$ forest, confirmed the $2\sigma$ confidence redshift range for the blazar $0.2315 < z < 0.322$. More recently, \citet{Dorigo22} improved the Ly $\alpha$ forest analysis and obtained the 95\% confidence redshift range of $0.227 < z < 0.254$. With such, the duration and the magnification of the {\em TESS} microlensing events in the S5\,0716+714 optical lightcurve, dictate that the lens is located in a close proximity to the source, $D_{\rm LS} \ll D_{\rm L}$, and in particular within the host galaxy of the blazar, as otherwise the implied lens parameters (mass and separation) would become unrealistic. Hence, below we take $D_{\rm S} \simeq D_{\rm L} \simeq d_{\rm A}$.

Moreover, the lens should be located within the central parts of the host, as otherwise the probability for the jet component to cross the folded caustic would be rather low, in contrary to the frequency of the microlensing events in the {\em TESS} lightcurve.

In general, S5\,0716+714 exhibits a stochastic variability at optical frequencies, with power spectral density (PSD) extending as a colored noise continuum over several decades of the variability timescales, from months and years down to hours and minutes \citep[see, e.g.,][]{Rani13,Bhatta16,Raiteri21}. In particular, the PSD analysis for the first season of the TESS campaign returns the best-fit power-law slope of $\alpha=2.01 \pm 0.04$ \citep{Raiteri21}, consistent with the overall red noise behavior. In addition, however, by means of analysing the source periodograms, the autocorrelation function, and the structure function, \citeauthor{Raiteri21} found the characteristic variability time-scales in the system of about about 1.7, 0.5, and 0.2\,day. It is not clear if these timescales could be linked to the micro-lensing events as discussed here, especially as there are indications that some of the rapid optical flares in the source covered with WEBT, are chromatic; yet 0.2 and 0.5\,day do correspond, at least roughly, with the duration of the flux drop/rise phases, and the total duration of the analyzed micro-lensing event candidates, respectively. Hence we cautiously conclude that the timing analysis of the TESS data do not rule out repetitive --- even in a quasi-periodical fashion --- microlensing-type structures in the red-noise lightcurve of S5\,0716+714.

The overall colored noise nature of the S5\,0716+714 optical variability, indicates that the observed emission of the blazar is produced by various emission components distributed within an extended segment of the jet, for which the scales range from tenths and hundredths of parsec down to the smallest scales set by the gravitational radius of the SMBH in the galactic nucleus,
\begin{equation}
    r_g = \frac{G M_{\bullet}}{c^2} \simeq \left(\frac{M_{\bullet}}{10^8 M_{\odot}}\right) \,{\rm AU}
\end{equation}
\citep[see in this context][]{Begelman08}. We posit that the microlensing events (as the ones analyzed here) become conspicuous whenever the innermost segments of the jet, with linear sizes $R\gtrsim r_g$, dominate the radiative output of the source. A very complex kinematics of the S5\,0716+714 jet --- with frequent changes in the jet position angle, as well as plethora of radio emission features on milliarcsec scales displaying various and often superluminal radial and non-radial motions \citep[e.g.,][]{Britzen10,Larionov13,Rani15} --- means that one should not hope to see any strict regularity in such microlensing events, as the source trajectory parameters $u_0$, $t_E$, and $\alpha$, may change/evolve with time. The encouraging finding, in this context, is that our fitting returns a rather narrow range for those parameters, in particular the Einstein crossing time each time of the similar order of a few/several days.

Keeping in mind the above considerations, for the analyzed microlensing events we therefore set the limit $R = \rho \, R_E \geq r_g$, which combined with Equation\,\ref{eq:angle} gives
\begin{equation}
    \left(\frac{M}{10^4\,M_{\odot}}\right) \geq \left(\frac{M_{\bullet}}{10^8 M_{\odot}}\right)^{2}\left(\frac{\rho}{0.01}\right)^{-2}\left(\frac{D_{\rm LS}}{100\,\textrm{pc}}\right)^{-1}.
\label{eq:mass1}
\end{equation}
On the other hand, since by definition $R_E = c \beta_{\rm app} \, t_E$, where $\beta_{\rm app}$ is the apparent velocity of the source emitting region through the lens, we also have
\begin{equation}
    \left(\frac{M}{10^4\,M_{\odot}}\right) \simeq 10^4 \,\left(\frac{\beta_{\rm app}}{30}\right)^2 \left(\frac{t_E}{2\,\textrm{d}}\right)^{2} \left(\frac{D_{\rm LS}}{100\,\textrm{pc}}\right)^{-1}  .
\label{eq:mass2}
\end{equation}
Unfortunately, the SMBH mass in S5\,0716+714 is unknown, and optical measurements of $\beta_{\rm app}$ are not available. However, for a blazar of the BL Lac type, we expect in general $M_{\bullet} \gtrsim 10^8 M_{\odot}$ \citep{Plotkin11}, and for S5\,0716+714 in particular $\beta_{\rm app} \lesssim 30$ has been measured at radio wavelengths \citep{Rani15}. With such, setting in addition $\rho = 0.01$, and also $t_E \simeq 2$\,day as the average crossing time (see Table\,\ref{tab:results}), the lower and upper limits are set by Equations\,\ref{eq:mass1} and \ref{eq:mass2}, respectively, correspond to the range of the binary lens mass $10^4 < M/M_{\odot} < 10^{8}$ for the distance of the lens from the active nucleus $D_{\rm LS} \simeq 100$\,pc, and $10^2 < M/M_{\odot} < 10^{6}$ for $D_{\rm LS} \simeq 10$\,kpc. We emphasize that the apparent velocity of the jet features inferred from radio observations with milli-arcsecond resolution may not necessarily provide a proper characterization of the innermost segments of the jet dominating the optical emission subjected to the micro-lensing events (as proposed here). Indeed, in the framework of the well-established MHD model for relativistic jets launched by accreting black holes, the jet bulk velocity increases from the jet base due to a gradual collimation and acceleration of the plasma outflow by magnetic forces \citep[e.g.,][and references therein]{Lyubarsky09}. Hence, apparent velocities measured in the radio domain at larger (parsec) distances from the center, $\beta_{\rm app} \lesssim 30$ in this case, provide us a very robust \emph{upper limit} for the proper motion velocity of the optical emission regions.

Another constraint on the lens could, in principle, follow from the requirement that the binary parameters should be stable during a single lensing event. However, the orbital velocity of the companion, 
\begin{equation}
    v_{\rm orb}  =  \sqrt{\frac{G M}{d \, R_E}} \sim  370 \left(\frac{M}{10^4\,M_{\odot}}\right)^{1/4}\!\!\left(\frac{D_{\rm LS}}{100\,\textrm{pc}}\right)^{-1/4}  \! {\rm km\,s^{-1}},
\label{eq:orbital}
\end{equation}
is sufficiently low for the range of $M$ and $D_{\rm LS}$ discussed above, that the distance traveled by the companion during the Einstein crossing time, is always less than the separation of the lens. In other words, the condition $v_{\rm orb} \, t_E \ll d \, R_E$ is easily satisfied for a wide range of lens masses located at larger distances, given the Einstein radius of the lens
\begin{equation}
    R_E \simeq 100 \, \left(\frac{M}{10^4\,M_{\odot}}\right)^{1/2} \left(\frac{D_{\rm LS}}{100\,\textrm{pc}}\right)^{1/2}  \, \textrm{AU} \, .
\label{eq:RE}
\end{equation}
On the other hand, the configuration of a binary lens could change over the $20$-yr timescale (since the WEBT 2003 event), especially for the lower ranges of the mass $M$ and the source-lens distance $D_{LS} < 100$\,pc. Changes in the position of the $0.00545\,M$ companion on the orbit around $M$, will lead to the rotation of the magnification pattern. But a double-horn feature in the lightcurve can be obtain even for $\alpha\simeq 90^{\circ}\pm 20^{\circ}$, at the expense of a slight asymmetry of the ``horns'' (i.e., of the magnified flux maxima). Moreover, the characteristic micro-lensing pattern can still be produced even with larger rotation angles, by means of more substantial changes in the source trajectory parameter, introduced by the precession of the jet on a similar timescale of years/decades; indeed, observational evidence for such a precession in S5\,0716+714 has been discussed in, e.g., \citet{Bach05} and \citet{Nesci05}. These two effects --- namely, the orbital modulation of the lens and the jet precession, both taking place on comparable timescales, leading altogether to a systematic evolution in the magnification patterns for the lensing events --- could, in fact, offer an exciting opportunity for an observational verification of the model proposed, if only the target could be monitored at optical and radio frequencies over a longer period of time with sufficiently dense coverage.

Yet an another consequence of the event detected by WEBT in 2003 is that one may require the Einstein radius to be large enough to cover the blazar over a timescale of, at least, $\tau \gtrsim 20$\,yr despite the host galaxy rotation. For a typical blazar with a $M_{\bullet} \sim 10^8 M_{\odot}$ black hole, the velocity dispersion of the host is expected to be around $\sigma_{\rm vel} \simeq 200$\,km\,s$^{-1}$. With such, assuming the lens is beyond the central SMBH sphere of influence (central $\sim 10$\,pc region), the condition $\sigma_{\rm vel} \, \tau < R_E$ gives 
\begin{equation}
    \left(\frac{M}{10^4\,M_{\odot}}\right) > 10^2\,\left(\frac{\sigma_{\rm vel}}{200\,\textrm{km/s}}\right)^{2}\,\left(\frac{\tau}{20\,\textrm{yr}}\right)^{2}\,\left(\frac{D_{\rm LS}}{100\,\textrm{pc}}\right)^{-1},
\label{eq:mass3}
\end{equation}
which provides the strongest lower limit on the mass of the lens. In particular, the above equation\,\ref{eq:mass3}, together with the upper limit following from equation\,\ref{eq:mass2}, yield $10^6 < M/M_{\odot} < 10^{8}$ for the distance $D_{\rm LS} \simeq 100$\,pc, and $10^4 < M/M_{\odot} < 10^{6}$ for $D_{\rm LS} \simeq 10$\,kpc. 

All in all, our analysis of the optical lightcurves of S5\,0716+714 resulting from the {\em TESS} observations, is consistent with the presence of a massive binary lens, containing an IMBH, possibly even a super-massive one, and a much less massive companion (by a factor of $\lesssim 0.01$), located within the host galaxy of the blazar (most likely the central kpc regions).

Until now, no robust detections of any radiative or dynamical signatures for the presence of IMBHs have been reported, although several models and scenarios are being discussed in the literature identifying IMBHs with ultraluminous X-ray sources, centers of dwarf galaxies, or centers of globular clusters \citep[for reviews see][]{Mezcua17,Greene20}. Interestingly, gravitational lensing of background stars by globular clusters in our Galaxy, has been proposed as a viable method for detecting such IMBHs \citep{Kains16,Kains18}. Globular clusters are moreover expected to be more numerous in massive ellipticals than in galaxies such our Milky Way \citep{Lim20,Harris17}.

We also note in this context recent N-body simulations of globular clusters, which not only reveal the presence of IMBHs, but also the fact that most of such IMBHs are in binary systems \citep[see, e.g.,][]{Konstantinidis13,Leigh13}. According to our modeling, the IMBH lens in the host galaxy of S5\,0716+714, should be accompanied by a much less massive (ratio 1:200) object. This mass ratio is within the range of the \textit{intermediate mass-ratio inspirals} (IMRIs), mergers of which can be probed in the near future directly with the \textit{Laser Interferometer Space Antenna} (LISA) gravitational wave detector \citep[see the recent review by][]{Amaro22}, with the conservative estimate of the detection rate of about two per year for the events at high redshifts \citep{Pestoni21}.

Currently, the only observational clue we have regarding the population of binary black holes approaching the IMBH range, is a large number of black hole binaries discovered by the Advanced LIGO and Advanced Virgo gravitational wave detectors, with the merger rate density estimated at $23.9^{+14.9}_{-8.6}$\,Gpc$^{-3}$\,yr$^{-1}$ \citep{Abbott21}, albeit in a lower mass range of $14-150\,M_{\odot}$. Until now, no signal for merging of more massive systems have been found in the LIGO/Virgo data \citep{Abbott22}, and the estimated rate of mergers for binaries with the total masses of $150\,M_{\odot}$ is $\sim 0.08$\,Gpc$^{-3}$\,yr$^{-1}$.

We propose that lensing events of the sort we advocate for S5\,0716+714, are in fact not uncommon among blazars, but the actual number of lensed sources, and lensing events in a single source, could be estimated only with a systematic optical monitoring of a larger number of targets with a sufficiently dense sampling. The underlying red noise-type intrinsic variability of blazar jets, is the other major obstacle in a robust identification of the micro-lensing events; finding a proper statistical test in this context is not straightforward, and relates to a wider open problem of assigning a significance to recurring features in the periodograms of colored noise time series. The events we have selected in the {\em TESS} lightcurve of S5\,0716+714, are all very regular, symmetric in shape, and in addition of a comparable duration. Moreover, the events are found during different flux levels of the source. While these improve our confidence in the lensing hypothesis, we caution the reader that the precise physical mechanisms that lead to intrinsic blazar optical variability have not yet been identified which makes it difficult to conclusively establish lensing as the cause of the symmetric variations with the available data.

Lastly, we emphasize again an important aspect of the proposed scenario, namely that we do expect the proposed lensing events in the optical lightcurve of S5\,0716+714 to repeat, even though not necessarily in a periodic or even quasi-periodic fashion. Moreover, the orbital modulation of the lens expected to take place on the timescale of years/decades for the lens-source distances $<100$\,pc and large lens masses $M>10^4M_{\odot}$, offers an exciting possibility for observational tests of the model by detecting a systematic evolution in the magnification patterns for the lensing events (modulo systematic changes in the jet trajectory parameters due to the jet precession).

\begin{acknowledgements}

D.~{\L}.~K. and {\L}.~S. were supported by the Polish NSC grant 2016/22/E/ST9/00061 and DEC-2019/35/O/ST9/04054. C.~C.~C. was supported by NASA DPR S-15633-Y. The authors thank M.~C.~Begelman and R.~D.~Blandford for their encouragement and comments, and the anonymous referee for their careful reading of the drafts, valuable suggestions, and an in-depth discussion.

\end{acknowledgements}

\bibliographystyle{aasjournal}

\end{document}